\documentclass[runningheads]{llncs}
\pdfoutput=1
\usepackage{xcolor}
\usepackage{graphicx}
\usepackage{leob}
\usepackage{booktabs}
\usepackage{caption}
\usepackage[tight,footnotesize]{subfigure}

\begin{document}


\title{StartupBR: Higher Education's Influence on Social Networks and Entrepreneurship in Brazil}
\titlerunning{Higher Education, Social Networks, and Entrepreneurship in Brazil}

\authorrunning{Michelle Reddy et al.}

 \institute{Graduate School of Education, Stanford University\\ \scriptsize{\email{mireddy@stanford.edu}} \and  \small{ Federal University of Technology - Paraná}\\ \scriptsize{\email{
julio.2018@alunos.utfpr.edu.br}~\email{thiagoh@utfpr.edu.br}}\and \small{Federal University of Minas Gerais} \\ \scriptsize{\email{yuri.pereira,~leob@dcc.ufmg.br}}  \and   \small{IBM Research}\\ \scriptsize{\email{marisaav@br.ibm.com}} \and \small{Computer Science Department, Stanford University}\\ \scriptsize{\email{leobo,~horowitz@stanford.edu}}}
   \author{Michelle Reddy\inst{1}, Júlio C. Nardelli\inst{2}, Yuri L.\ Pereira\inst{3},  \\ Marisa Vasconcelos\inst{4}, Thiago H. Silva\inst{2}, Leonardo B.\ Oliveira
\inst{3,5}, Mark Horowitz\inst{5}}

\maketitle              %
\begin{abstract}
Developing and middle-income countries increasingly emphasize higher education and entrepreneurship in their long-term development strategy. Our work focuses on the influence of higher education institutions (HEIs) on startup ecosystems in Brazil, an emerging economy. First, we describe regional variability in entrepreneurial network characteristics. Then we examine the influence of elite HEIs in economic hubs on entrepreneur networks. Second, we investigate the influence of the academic trajectories of startup founders, including their courses of study and HEIs of origin, on the fundraising capacity of startups. Given the growing capability of social media databases such as Crunchbase and LinkedIn to provide startup and individual-level data, we draw on computational methods to mine data for social network analysis. We find that HEI quality and the maturity of the ecosystem influence startup success. Our network analysis illustrates that elite HEIs have powerful influences on local entrepreneur ecosystems. Surprisingly, while the most nationally prestigious HEIs in the South and Southeast have the longest geographical reach, their network influence still remains local. 

\end{abstract}
\section{Introduction}\label{sec:intro}
\setlength{\parskip}{0pt}

Entrepreneurship and higher education are increasingly viewed as drivers for long-term sustainable economic development. Despite this strong policy focus, studies of entrepreneur networks focus exclusively on high-income countries.  For resource-rich and emerging economies like Brazil transitioning to a more sustainable, knowledge-based economy, higher education, and entrepreneurship are particularly important. At the same time, many resource-rich and emerging economies have high levels of inequality.  Brazil is characterized by spatial inequalities, most notably, among regions, and evident in the stark contrast between the Brazilian North and Northeast and the economic hub of the South and Southeast, despite improvements in recent years \cite{Silva2017}. Educational inequality in particular is well-documented, particularly in Brazil \cite{Sanderson2017} and across emerging economies \cite{Balestra2018}, and globally in terms of higher education access \cite{McCowan2007,UNESCO2016,Msigwa2016}.  

If entrepreneurship is heralded as the pathway towards sustainable development, to what extent do Higher Education Institutions (HEIs) influence entrepreneur networks? Will elite HEIs perpetuate existing inequalities, particularly across regions, by having more influence on startup ecosystems? Using social network analysis and mining public Web data of Brazilian entrepreneurs, we hypothesize that entrepreneur networks in regionally disadvantaged areas, such as the Brazilian Northeast, are closely linked with networks from top universities in the wealthier Southeast. We also conceive that the nature of networks will vary by region, given their varying levels of development, as some regions have more access to capital and others to natural resources. In addition, we test our assumption that at the regional level, elite HEIs will influence startups through the social networks formed through HEIs. Notably, we discuss how elite HEIs, according to national educational quality rankings, drive the success of a regional entrepreneurial ecosystem. Overall, we investigate the nature of these networks within Brazil and examine how universities contribute to Brazil’s regional entrepreneur networks. Specifically, we aim to address the following questions:  

i) To what extent do HEIs influence entrepreneurial networks, within and outside their region? 
ii) How do entrepreneur networks vary by region in Brazil?
iii) Are entrepreneur networks mostly embedded in elite HEI networks?  

As networks provide entrepreneurs with information, capital, and services, we examine entrepreneur networks in Latin America’s largest startup ecosystem, Brazil, in this study \cite{Lechner2003,Renzulli2005}. We chose Brazil because there is limited, if any, empirical analysis of the conditions fostering high-tech regions in middle-income countries. In particular, Brazil is a suitable case because while specific regions are middle-income, others, such as the Northeast, have GDPs similar to low-income countries. Given these stark contrasts, we look at HEIs and their influence on regional entrepreneur networks.  Through our Brazil analysis, we explore HEI influence on high-tech ecosystems in both a middle and low-income context. 

We use computational methods to mine public data from an online database regarding entrepreneurs in Brazil and triangulate with information publicly available in a social media network, as well as with official open data from Brazil's Ministry of Education. First, we download data regarding our target Brazilian startup ecosystems from Crunchbase \cite{CrunchBase} database. Second, we collect relevant data from LinkedIn to enrich our initial data on startup ecosystems. Finally, we add information about the General Index of Courses, which is an official indicator of quality concerning HEIs in Brazil. Note that the use of computational methods here is especially important since Brazil is a continental country and conventional data collection methods like questionnaires and interviews do not scale well.  First, we characterize Brazil's entrepreneur network at the national level. Then, we create a framework for investigating the influence of HEIs, in particular, elite HEIs, degree programs, and educational quality, on entrepreneur networks. Overall, our study contributes to education, entrepreneurship, and development research. 

\section{Related Work}\label{sec:rw}

The {\em entrepreneurial university} is a global phenomenon, due to the internal development of the university \cite{ETZKOWITZ2000313} and as the transition to a knowledge-based economy became a goal for sustainable economic development \cite{LABRA201678}. Entrepreneurial activities enhance national and regional economic growth as well as university finances \cite{ETZKOWITZ2000313}. Just as Brazil made strides in terms of startup growth in the past decade, it has exponentially increased access to higher education. Yet, inequalities still remain. Other resource-rich countries, such as Qatar \cite{Gremm2018}, Malaysia, and Saudi Arabia \cite{Kumar2013}, increasingly invest in higher education to move towards a knowledge-based economy and away from natural resource dependency.  

The influence of universities on sciences and technology-based industries is well-documented (see for example \cite{Rosenberg1994}).  University-industry linkages include the movement of university graduates into commercial firms and faculty entrepreneurship, faculty involvement on advisory boards, industry gifts supporting university research and student training, among others \cite{porter2005institutional}. There is a tendency for the research and development efforts of organizations to spillover into the innovation efforts of other organizations \cite{Jaffe1986}, which can occur across industries but is particularly acute within regions, and amplified when key participants are research organizations \cite{Dasgupta1994,Owen2004}. In particular, HEIs, and their relationship with industry, may be more favorable in certain regions than in others \cite{porter2005institutional}, especially if more elite universities are clustered in economically wealthy regions.

The technological revolution enabled new entrepreneurial initiatives worldwide, creating an enabling environment for business without the startup costs of the larger firms that dominated the economic landscape of the mid-twentieth century in developed countries \cite{Comission2013}. While technology is vital in the rise of entrepreneurship worldwide, as Banerji and Reimber \cite{BANERJI201946} note, the importance of social networks on entrepreneurship is intuitive. In particular,  potential funding agencies predict startup success by examining the social networks of founders \cite{BANERJI201946}, and networks provide information and opportunities \cite{BURT2000345,LARSON1991173}, and legitimacy \cite{Klyver2008}.

The role of social ties in entrepreneur networks has also been observed by numerous studies. In particular, Zimmer and Aldrich \cite{Zimmer1987} note the importance of social networks on all three aspects of entrepreneurial success: launching a startup, turnover, and sustainability. These findings hold across several cultural contexts, for instance, in China \cite{BATES1997109}, as well as for ethnic minorities in the United States \cite{Light1984}. Therefore, our study results are potentially useful for other cultural contexts, and in particular, middle-income and developing countries.


\section{Data and Methods} \label{sec:data}

\subsection{Overview}\label{sec:SecData}

We explore three datasets in this study, namely:

\textbf{Crunchbase.} Crunchbase is a global database updated daily that contains information about companies, funders, and staff \cite{crunch12,Dalle2017}. As a partially crowd-sourced database, Crunchbase is increasingly used for academic and commercial purposes  \cite{Dalle2017}. We acquired a commercial license enabling unlimited access in addition to advanced search functions in Crunchbase. We procured all available data of Brazilian companies up to August 26, 2018. For 3,375 companies throughout Brazil, we include company name; LinkedIn profile URL; founding date; company type (or category); the total of investments received; and headquarters location. As Crunchbase uniquely links other data sources such as Twitter \cite{Tata2016,TATA201738} and LinkedIn \cite{Nuscheler2016,Dalle2017}, we linked Crunchbase with LinkedIn to examine characteristics of startup founders, their universities, and social networks.

 \textbf{LinkedIn.} LinkedIn, as a popular social network of professional contacts, provided the educational information of the company founders. We collected the profiles of employees that held titles such as CEO, owner, and founder with the LinkedIn profile URL obtained through Crunchbase. In the end, this yielded 1,177 profiles and the main data collected were: degree type/level (e.g., Bachelor, Master or Ph.D.); degree area (e.g., Sociology, or Computer Science); graduation year; and the name of the alma mater. Multiple degrees for the same profile were common. We collected all information on the LinkedIn profiles.
 
\textbf{IGC.} The General Index of Courses (IGC\cite{IGC,OECD}) is the official quality indicator for HEIs in Brazil. Annually, the National Institute of Educational Studies and Research (INEP\cite{INEP,OECD}) performs the Census of Higher Education (CENSUP\cite{CENSUP}), which is used to calculate the IGC, a metric of HEI quality. We used the IGC to classify HEIs as {\em elite} or {\em nonelite} instituions\footnote{Stanford and USP are absent from IGC rank. Yet, due to their academic excellence \cite{Rank}, we regarded them as elite HEIs.}.

\subsection{Data Pre-Processing}\label{sec:DataPropro}

For data pre-processing,  we first obtained the geolocation of companies and HEI addresses. We used Google Geocode API to yield formatted address and geographic coordinates (e.g., latitude and longitude). We also standardized the name field. As Crunchbase has over 1,400 different categories for companies, we matched Crunchbase categories to categories used by the Brazilian Association of Startups (Abstartups)~\cite{Xray2017}. Since LinkedIn users report their educational background by open response, we standardized the names of HEIs using the IGC list, from the INEP website\cite{CensoHEI}, and searched phonetically, using manual coding when necessary, to match IGC and Linkedin HEI names.

Here, we describe our methodological approach to identifying startup ecosystems. Most studies of entrepreneur networks are rich in interview and survey data \cite{BANERJI201946}, see, for example, Zimmer and Aldrich (1987)\cite{Zimmer1987}, Bates (1987)\cite{BATES1997109}, and Light (1984)\cite{Light1984}. Recent access to databases such as LinkedIn and Crunchbase facilitates more generalizable results, given the ability to generate a larger sample size \cite{BANERJI201946}. Thus, we draw on LinkedIn and Crunchbase databases and use social network analysis to test our main research questions. 

We considered {\em startups} companies that are at most 15 years old.  From the $3,375$  companies we extracted from Crunchbase, we selected only $1,957$ (57.98\%). Next, we grouped the startups by city and considered those cities with at least ten startups as {\em ecosystems}. Then, we examined only startups associated with our ecosystems. As a result, we got 21 ecosystems covering  $1,547$ startups (45.83\%) of our initial set. We then collected founders' data from LinkedIn, yielding 146 HEIs and 648 academic degrees of founders. (\strt{Overview} summarizes our dataset numbers.) \strf{MapEcosystems} shows the geographical distribution of the ecosystems present in our dataset.

\begin{figure}[!tbh]
    \centering
    \begin{minipage}{.45\textwidth}
       \includegraphics[width=\linewidth]{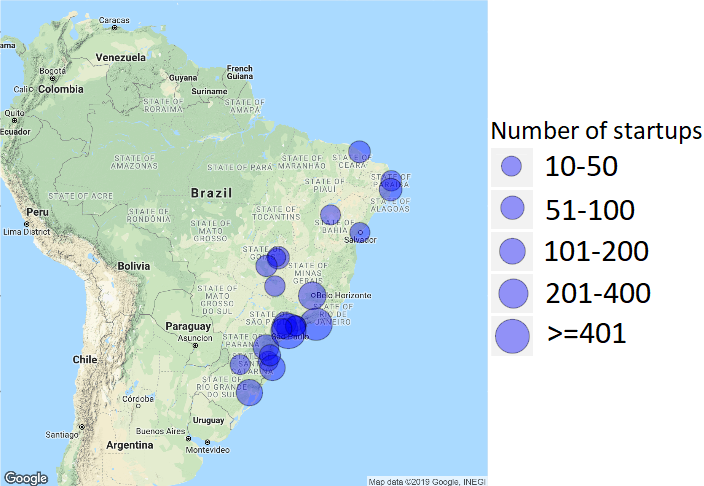}
        \caption{Map of Brazilian Ecosystems. The redder and larger the circle, the greater the number of startups. The center of the circle indicates the location of the ecosystem.}
        \label{fig:MapEcosystems} 
    \end{minipage}%
     \hfil
    \begin{minipage}{0.45\textwidth}
     \includegraphics[width=\linewidth]{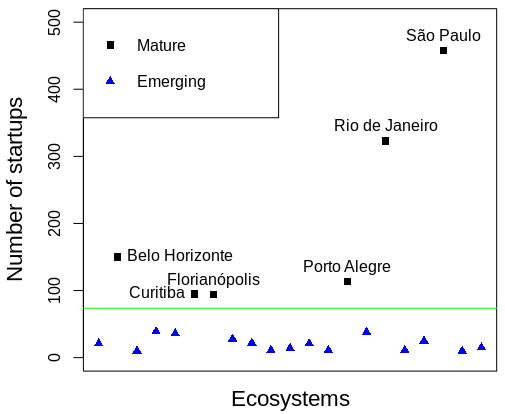}
        \caption{Brazilian Ecosystems by City. {\em Mature} ecosystems are in the cities where the number of startups is greater than the national mean (green line), in contrast to {\em emerging} ecosystems.}
        \label{fig:EcosystemOutlier}
       
    \end{minipage}
\end{figure}

To address the gap regarding the number of startups among Brazilian systems, we divided the ecosystems into {\em mature} and {\em emerging} ecosystems. \strf{EcosystemOutlier} illustrates the difference between the two groups. Ecosystems with 74 startups or more are considered mature ecosystems. According to Crunchbase, the largest ecosystems (\strt{Developed}) are in Brazilian state capitals such as São Paulo (SP); Rio de Janeiro (RJ); Belo Horizonte (MG); Porto Alegre (RS); Curitiba (PR); and Florianópolis (SC). Together, they comprise 79.82\% of startups and 97.05\% of total fundraising. All of the largest ecosystems are located in the South or Southeast, the economic hub of Brazil. The emerging ecosystems, on the other hand, encompass $15$ cities (\strt{Emerging}). Emerging ecosystem locations are more diverse in terms of region and city size, ranging from regional capitals like Brasilia (DF), Fortaleza (CE), and Goiânia (GO) to smaller cities like Uberlândia (MG), Joinville (SC), and São José dos Campos (SP).

\begin{table}[!tbh]
    \centering
\begin{minipage}{0.46\textwidth}
\scriptsize
\centering
\caption{Mature startup ecosystems.}
\begin{tabular}{llrr}
\hline
\textbf{Ecosystem} & \textbf{Region} & \textbf{size\footnote{{\em size} stands for the number of startups.}} & \textbf{Fundraising}  \\
\hline
São Paulo           & Southeast & 458   & \$2.6B    \\
Rio de Janeiro      & Southeast & 323   & \$283.6M  \\
Belo Horizonte      & Southeast & 151   & \$33.8M   \\
Porto Alegre        & South     & 114   & \$11.1M   \\
Curitiba            & South     & 95    & \$67.3M   \\
Florianópolis       & South     & 94    & \$43.5M   \\  
\end{tabular}
\label{tab:Developed}
    \end{minipage}
    \hfil
    \begin{minipage}{.46\textwidth}
\scriptsize
\centering
\caption{Top emerging ecosystems.}
\begin{tabular}{llrr}
\hline
\textbf{Ecosystem}   & \textbf{Region} & \textbf{size} & \textbf{Fundraising}  \\ 
\hline
Brasilia               & Central-West  & 39    & \$3.7M    \\
Recife                & Northeast     & 38    & \$8.5M    \\
Campinas               & Southeast     & 36    & \$13.4M   \\
Fortaleza              & Northeast     & 28    & \$105.4K  \\
SJ Campos        & Southeast     & 25    & \$4.1M    \\
Goiânia                & Central-West  & 22    & no info.  \\
Barueri                & Southeast     & 21    & \$10.9M   \\
Joinville             & South         & 21    & \$41.3M   \\
Uberlândia             & Southeast     & 15    & \$2.6M    \\
João Pessoa            & Northeast     & 14    & \$194.8K  \\
\end{tabular}
\label{tab:Emerging}
    \end{minipage}
\end{table}

\section{Results}
\label{sec:res}

\strs{network_char} explores the network relationship between startups and HEIs, and the academic trajectory of startup founders. \strs{Net} analyzes the relationship between startups and HEIs. Finally, we investigate how educational quality influences the success of an ecosystem in \strs{EliteHEI}.

\subsection{Network Characterization}
\label{sec:network_char}
Here we analyze founders in terms of HEI,  major, and degree nature (type and level). \strf{Scholarity} shows the degrees held by company founders before and after company creation. The most common degree is the Bachelor's degree, followed by the MBA and then other master's degrees. 
We find that founders obtain most of their Bachelor degrees before startup creation (\strf{Scholarity}, left). Besides, after startup creation (\strf{Scholarity}, right), the demand for other courses increased $50$\% (Master), $23$\% (MBA),  $156$\% (Extension\footnote{In Brazil, {\em extension} courses are certified programs that do not require a Bachelor's degree, like {\em continuing studies} in the U.S.}), and $647$\% (Ph.D.). 
This suggests that, after launching a startup, some founders may look for new educational opportunities that may add value to their business.

\begin{figure}[!tbh]
    \centering
    \begin{minipage}{0.40\textwidth}
        \includegraphics[width=\linewidth]{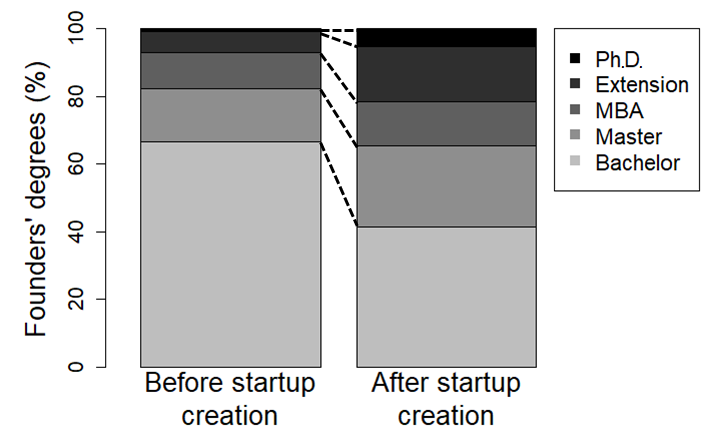}
        \caption{Founders' degrees.}
        \label{fig:Scholarity}
    \end{minipage}
    \hfil
    \begin{minipage}{.55\textwidth}
        \includegraphics[width=\linewidth]{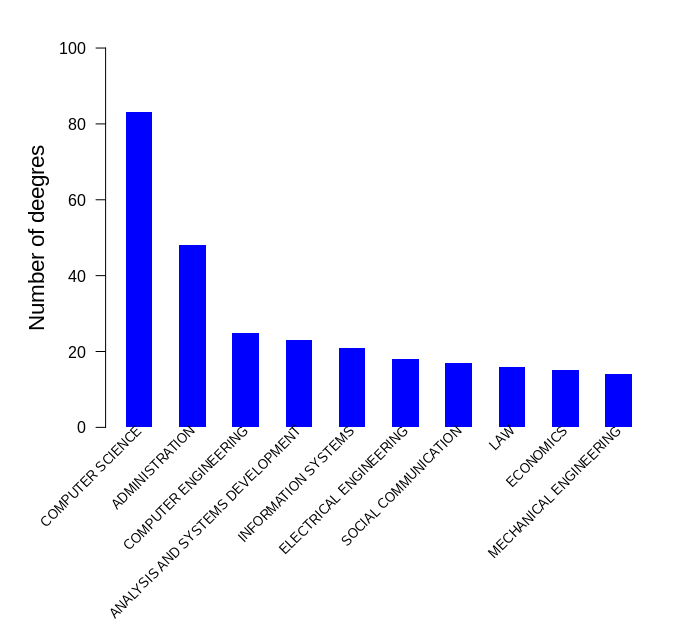}
        \caption{Popular majors pre-startup creation.}
        \label{fig:TopBachelors}
    \end{minipage}
\end{figure}

\strf{TopBachelors} presents the  Bachelor's degree courses taken by the founders before startup creation. Most degrees come from STEM (Science, Technology, Engineering, and Mathematics) ($\approx$59\%) and social sciences ($\approx$39\%). Computer Science is the most popular course of study among startup founders, and many other courses are related to Computer Science (e.g., Computer engineering). Nearly half of startups are in IT or Telecom, perhaps drawing on the computer science background of many founders (\strf{CatStartups}).

In a next step, we also examine whether founders of the same startup have similar academic trajectories. For each founder, we consider an academic trajectory vector where the $i$th position represents the number of degrees concluded in HEI $i$. We then measure the academic trajectory similarity between each pair of founders of the same startup using cosine similarity, then we average those values aggregating by startup. \strf{SimilarityIES} shows the cumulative distribution function (CDF) of this average similarity coefficient.

Note that approximately 55\% of the startups have non-zero cosine similarity, which means that founders had at least one common HEI in their academic trajectory. 
By further investigating the data, we found that, for the same group startups, 83\% of them have contemporary founders  (i.e., studied at the same HEI during the same period). Many founders may have met while at university, through acquaintances, or other university affiliations, such as being in the same social network even after university.


\begin{figure}[!tbh]
    \centering
 \begin{minipage}{0.54\textwidth}
    \includegraphics[width=\linewidth]{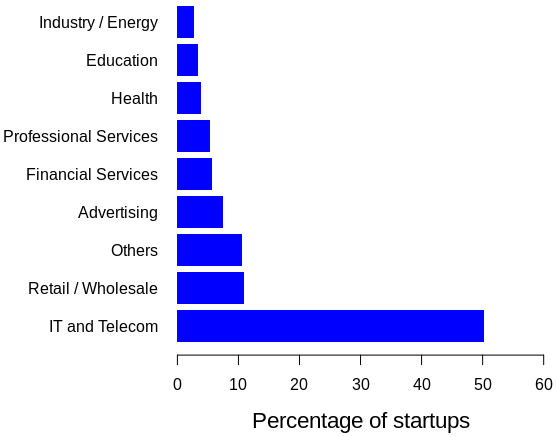}
    \caption{Startup categories.}
    \label{fig:CatStartups}
\end{minipage}
     \hfil
          \begin{minipage}{.36\textwidth}
        \includegraphics[width=\linewidth]{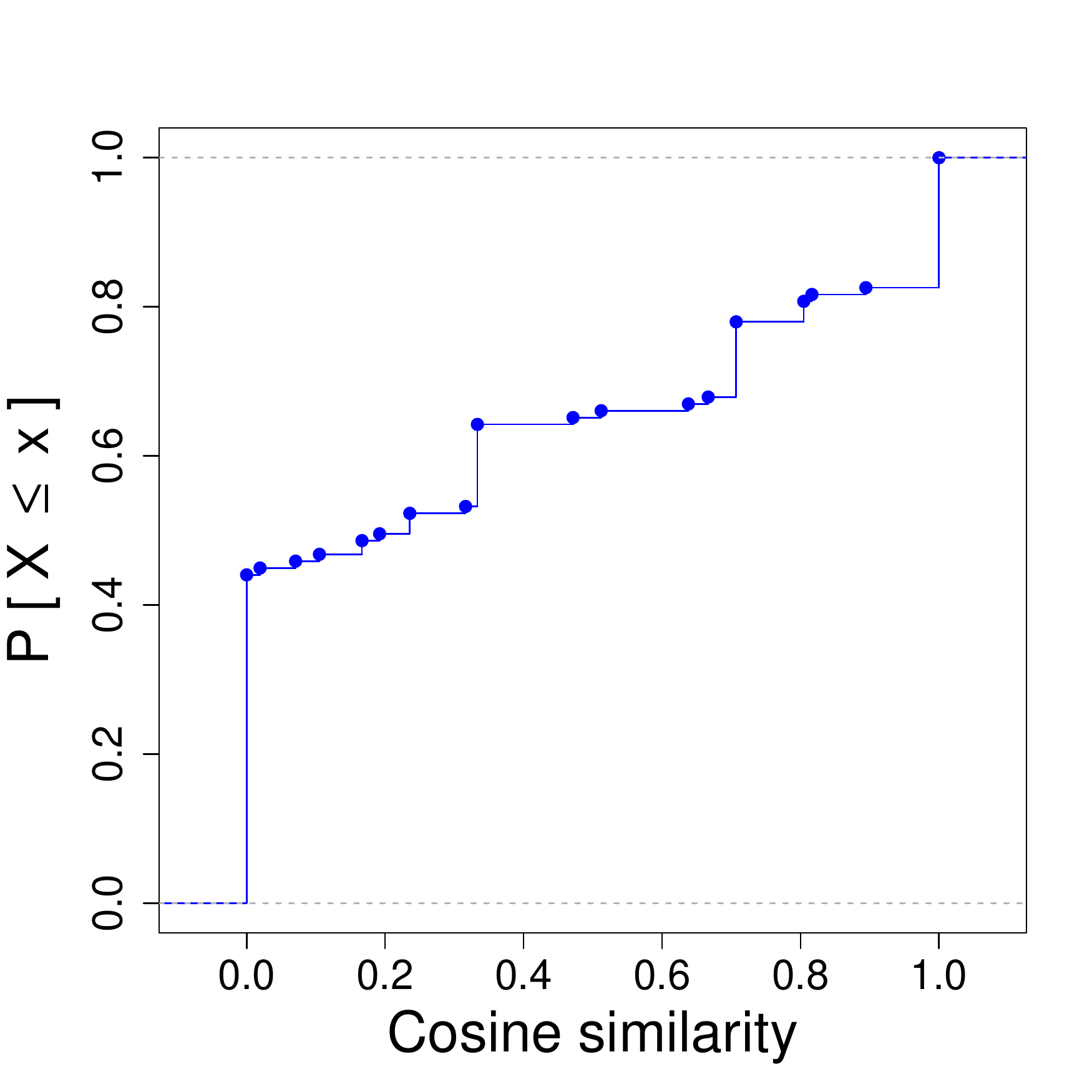}
        \caption{CDF of cosine similarity over HEI.}
        \label{fig:SimilarityIES}
    \end{minipage}%
\end{figure}


   

\subsection{Relationship Between HEIs and Startups}
\label{sec:Net}
In this section, we analyze the network relationship between HEIs and startups. Using social network analysis, we compare the described ecosystems in terms of academic trajectories, connectivity, spatial distribution. Finally, \strs{EliteHEI} analyzes the success of ecosystems as a function of HEI quality rankings.


\subsubsection{Network Approach}
\label{sec:Creation}


We use an undirected bipartite graph $G=(U,V,E)$, where nodes $v_i \in V$ are startups, nodes $u_i \in U$ are HEIs, and an edge $e_{i,j} =(v_i,u_j)$ exists from node $v_i$ to $u_j$ if a startup $v_i$ founder is a HEI $u_j$ alum. For our analysis, we consider two networks of this kind: (i) {\tt Undergrad} comprising only Bachelor degrees of founders; and (ii) {\tt All-Degrees} including any founder degree (Appendix~\ref{sec:app-nets}, \strf{NetUndergradStartupsAcdemic} and \strf{NetAllStartupsAcdemic}, respectively). Both networks also include the HEIs that issued the degrees.

\subsubsection{HEIs Centrality} \label{sec:centrlityTrad}
 \strt{CentraTypeDegree} shows the top ten HEIs according to networks'degree, closeness, and betweenness centralities \cite{Newman2010}.

Degree centrality reflects the importance of a node through its number of connections. Notably, in our network, HEIs are only linked to startups. So, the degree centrality expŕess the direct influence of the HEI in the startup formation. We found that University of Sao Paulo (USP) is the most central node in both {\tt Undergrad} and {\tt All-Degrees} networks (\strt{CentraTypeDegree}). In addition, an international HEI figures top-ranked on {\tt All-Degrees}: Stanford University. Upon closer examination, we find that these founders took extension courses at Stanford.


Broadly, closeness centrality captures the distance to all other nodes in the network. Here, the closeness centrality suggests that more elite HEIs reach (or influence) the network faster. 
In terms of undergraduate degrees among founders, Universidade Estadual Paulista (UNESP), though not top-ranked according to degree centrality, appears in the 2nd position in terms of closeness centrality, likely because UNESP is present in $24$ cities. Additionally, among the top-ranked HEIs in terms of founder undergraduate degrees, there is AIEC/FAAB\cite{AIEC}, a HEI that offers online courses nationwide. Finally, FGV/SP, Stanford, and IBMEC are the most central HEIs among {\tt All-Degrees}. This is likely due to their online course delivery and the high ranking of their business programs. 
 
Betweenness centrality tells how often a node is within the shortest path with another in the network.  In our study, this metric unveils HEIs that connect distinct social circles and then foster entrepreneurship. Here, Universidade Federal de Santa Catarina (UFSC) is the most central in {\tt Undergrad}, and Federal University of the State of Rio de Janeiro (UNIRIO) in {\tt All-Degrees} (\strt{CentraTypeDegree}).  Finally, centrality top-ranked HEIs, in general, are most elite (IGC $\geqslant 4$) HEIs. (There are two exceptions whose IGC $= 3$, though: AIEC and FDMC.) Also, 95 HEIs of 146 are located in major cities in the South or Southeast of Brazil, the economic hub of the country.




\begin{table}[thb]
\centering
\caption{Top 10 HEIs per degree, closeness, and betweenness centrality.}
\scriptsize
\begin{tabular}{cllllll}
\cline{2-7}
\multicolumn{1}{l}{}                  & \multicolumn{2}{c}{\textbf{Degree}}                                                & \multicolumn{2}{c}{\textbf{Closeness}}                                             & \multicolumn{2}{c}{\textbf{Betweenness}}                                          \\
\multicolumn{1}{l}{\textbf{Ranking}} & \multicolumn{1}{l}{\textbf{Undergrad}} & \multicolumn{1}{l}{\textbf{All-Degrees}} & \multicolumn{1}{l}{\textbf{Undergrad}} & \multicolumn{1}{l}{\textbf{All Degrees}} & \multicolumn{1}{l}{\textbf{Undergrad}} & \multicolumn{1}{c}{\textbf{All Degrees}} \\ \hline
\multicolumn{1}{c|}{\textbf{01}}      & USP                                     & USP                                       & FGV/SP                                  & FGV/SP                                    & UFSC                                    & UNIRIO                                   \\
\multicolumn{1}{c|}{\textbf{02}}      & UFRGS                                   & PUC/SP                                     & UNESP                                   & STANFORD                                  & FGV/SP                                  & FGV/SP                                   \\
\multicolumn{1}{c|}{\textbf{03}}      & UFRJ                                    & UFRGS                                     & USP                                     & IBMEC                                     & UNESP                                   & PUC/SP                                    \\
\multicolumn{1}{c|}{\textbf{04}}      & PUC/SP                                   & FGV/SP                                    & UFRJ                                    & UFMG                                      & USP                                     & STANFORD                                 \\
\multicolumn{1}{c|}{\textbf{05}}      & UFSC                                    & UFRJ                                      & UAM                                     & UFRJ                                      & UAM                                     & USP                                      \\
\multicolumn{1}{c|}{\textbf{06}}      & UFMG                                    & UFSC                                      & MACKENZIE                               & PUC/SP                                     & PUC/RS                                   & UFRJ                                     \\
\multicolumn{1}{c|}{\textbf{07}}      & PUC/RS                                   & UFMG                                      & UFMG                                    & USP                                       & UFRJ                                    & UFRGS                                    \\
\multicolumn{1}{c|}{\textbf{08}}      & PUC/PR                                   & IBMEC                                     & FDMC                                    & INSPER                                    & PUC/SP                                   & UFSC                                     \\
\multicolumn{1}{c|}{\textbf{09}}      & PUC/MG                               & STANFORD                                  & AIEC/FAAB                             & MACKENZIE                                 & UFRGS                                   & UFMG                                     \\
\multicolumn{1}{c|}{\textbf{10}}      & FGV/SP                                  & PUC/PR                                     & PUC/SP                                   & UAM                                       & UFMG                                    & IBMEC                                   
\end{tabular}
\label{tab:CentraTypeDegree}
\end{table}

\subsubsection{HEIs Spatial Degree Centrality}\label{sec:spatialCentr}

We draw on definitions by Lima and Musolesi \cite{Lima2012} for our spatial degree analysis below. Each node $i$,  $i \in V$ or $i \in U$ in our affiliation network $G=(U,V,E)$, is assigned a set of $j$ neighbours nodes, i.e., the neighbours of node $i$ is the set of nodes $j$ that are reachable from $i$ through the out-link $e_{ij}\in E$. All of them are represented by points on Earth $P_i = \{p_0^{(i)}, p_1^{(i)},...,p_{|j|}^{(i)} \}$, expressed through latitude and longitude.

For the spatial degree analysis, we first define a spatial neighborhood $S$ as a circular region specified by its center and radius. Given a node $i$, its spatial coordinates (latitude and longitude) represents the center of a spatial neighborhood $S_i$ with a certain radius. The intersection $P_i \cap S_i$ contains all the points, i.e., nodes representing HEIs and startups, falling inside the region $S_i$ that are neighbors of $i$ in $G$. In this way, we can compute the spatial degree centrality $C$ of node $i$ with spatial neighborhood $S$ as:

\begin{equation}
    C_{i,S} = |P_i \cap S_i|
\end{equation}

In this study, we are interested in the average spatial degree centrality $\overline{C}$ for HEIs. Thus, for our network $G=(U,V,E)$ this metric is expressed as:

\begin{equation}
   \overline{C}_{U,S} = \frac{1}{|U|} \sum_{u \in U} C_{u,S},
\end{equation}

where the set $U$ represent HEIs. 

\begin{figure}[tbh]
    \centering
    \begin{minipage}{.45\textwidth}
        \includegraphics[width=\linewidth]{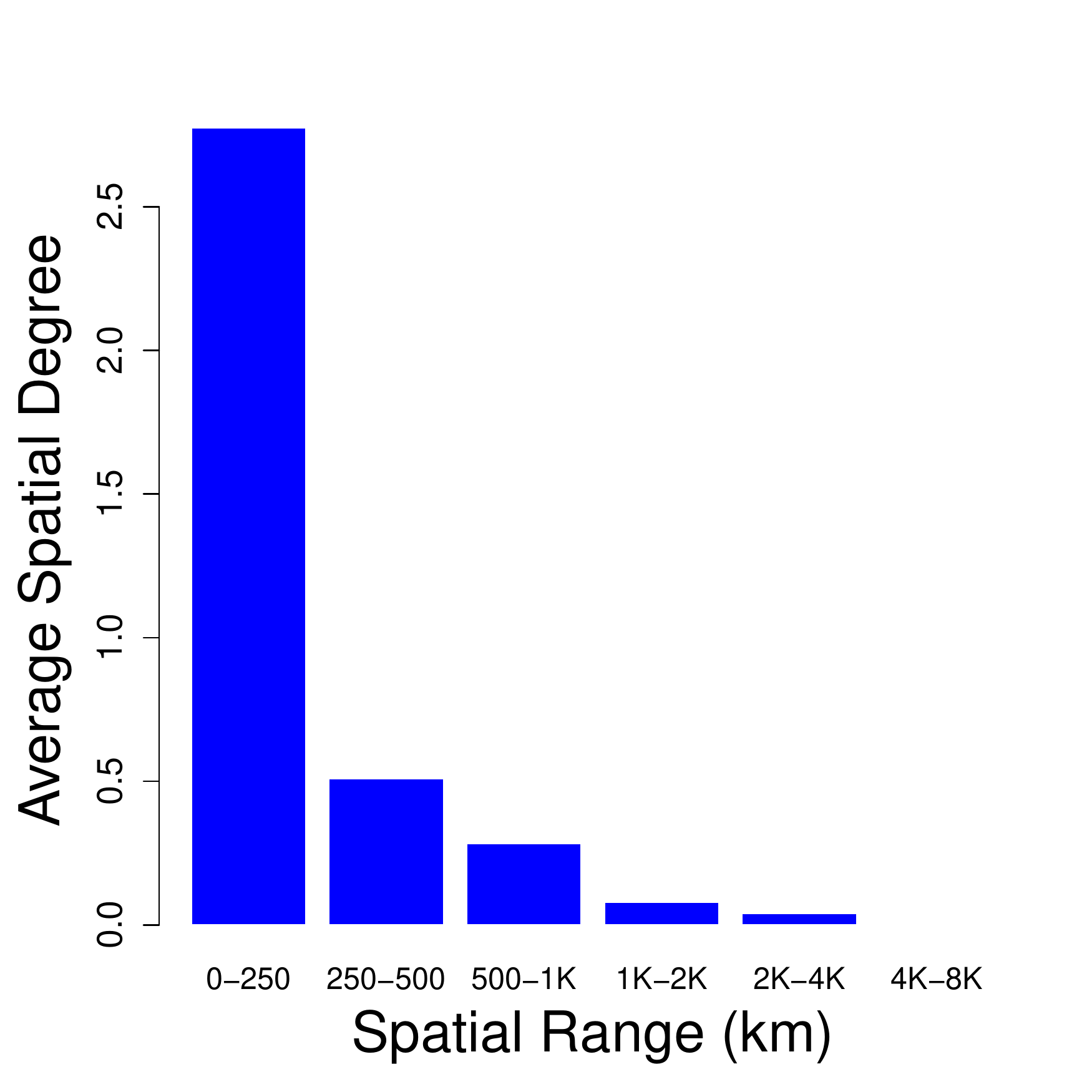}
        \caption{Spatial degree analysis for different spatial neighborhood $S$.}
        \label{figSpatialDegree}
    \end{minipage}%
     \hfil
    \begin{minipage}{0.40\textwidth}
        \includegraphics[width=\linewidth]{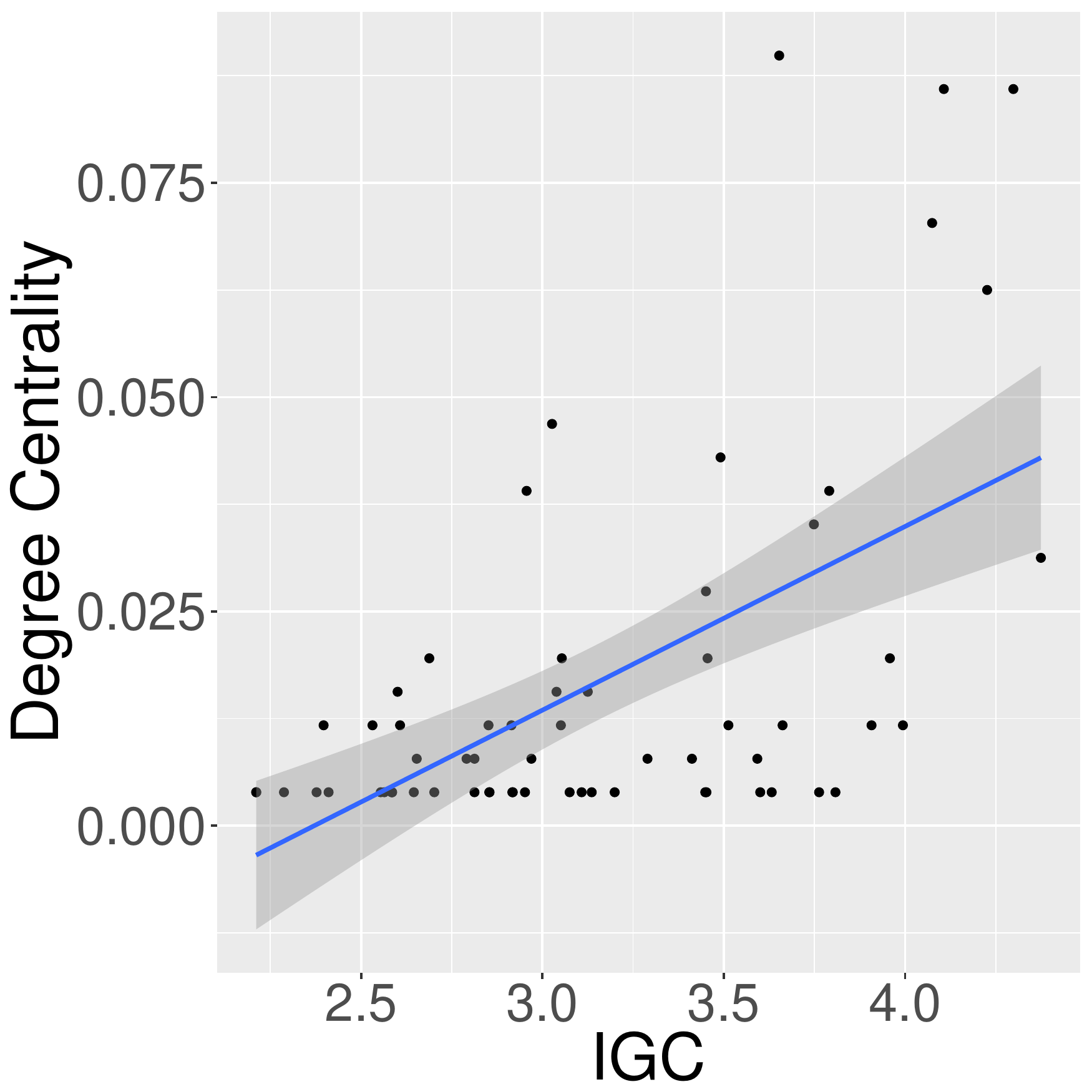}
        \caption{Scatterplot of node degree centrality and IGC. Pearson correlation is $0.56$ (p-value $<$ $0.001$).}
        \label{figCorrDegreeIGC}
    \end{minipage}
\end{figure}

Figure \ref{figSpatialDegree} shows the spatial degree considering different non-overlapping spatial ranges, meaning the ranges are circle and expanding annular rings around each HEI. This analysis takes into account a network composed of startups that founders obtained any degree from any HEI in the 15 years before the startup creation. Most of the connections are short distance, up to 250 km, suggesting that the influence of HEIs are mostly local. However, we find that elite HEIs, such as PUC/SP, UNICAMP, and IBMEC, have the longest spatial ranges, therefore, their influence is more likely to extend beyond their local ecosystem and into other regions.

In addition, we calculate the similarity of connections in the network concerning the nodes' state, using the assortativity coefficient \cite{newman2003mixing}. In general, the coefficient lies between $-1$ and $1$.  The network has perfect assortative mixing patterns when the assortativity coefficient is $1$,. The assortativity coefficient by {\em state} is $0.72$ for the same network studied in the spatial analysis. This means that the majority of connections happen between nodes from the same state, corroborating with what is observed by the spatial degree centrality analysis. Thus, HEIs, in general, have more influence within their region. 



\subsection{Elite HEI Alumni and Enhanced Startup Fundraising Capabilities}
\label{sec:EliteHEI}

We also examine how education quality drives the success of an ecosystem. We assume that HEIs with higher educational quality rankings are perceived as more elite. We calculated the Pearson correlation between the HEIs' IGC and the HEIs' degree centrality, and construct a scatter plot in Figure~\ref{figCorrDegreeIGC}. The Pearson correlation is moderate and around $0.56$ (p-value < $0.001$). Figure~\ref{figCorrDegreeIGC} also plots the linear regression, with a 95\% confidence interval of the best-fit line. These findings suggests that elite HEIs (high ranked IGC) have more startup connections, and overall support our hypothesis that elite HEIs have more influence on startup ecoystems.

We also analyzed the fundraising capability ($\kappa$) of startups. Equation \ref{eqP} describes how $\kappa$ is calculated for a given startup $i$:

\begin{equation}
    \kappa_i = \frac{F_{i,t_o}}{Li \times E_{i,t_o}},
    \label{eqP}
\end{equation}

where, $\kappa_i$  is the fundraising capability for startup $i$ up to time $t_o$ (now), $F_i$  the total fund raised in the life cycle (from creation up to $t_o$), $L_i$ , startup age in months, and $E_i$  current number of employees. This equation was also used by Perotti and Yu \cite{Perotti2015}. 

\begin{figure}
    \centering
    \includegraphics[width=0.45\linewidth]{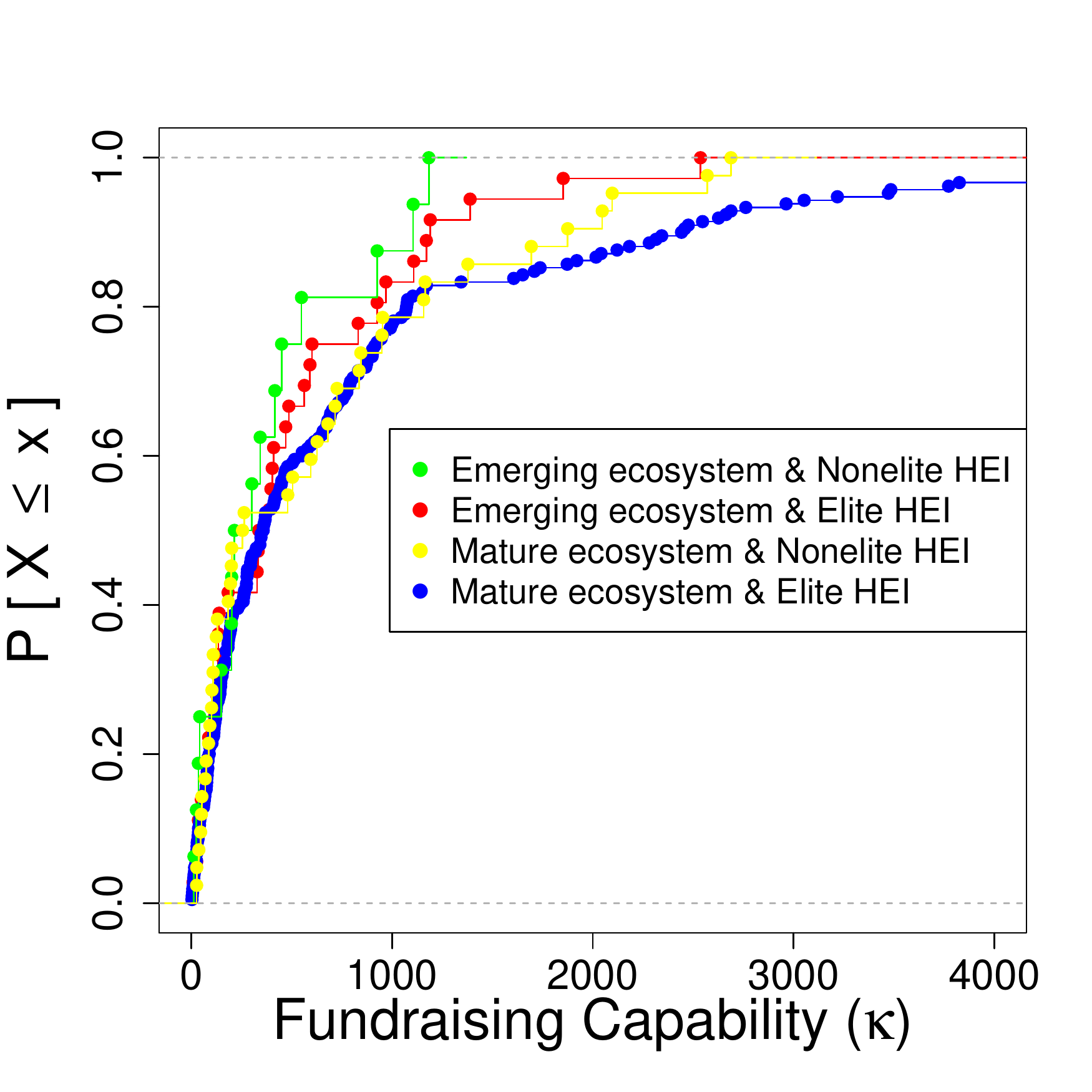}
        \caption{CDF for fundraising capability ($\kappa$) of startups.}
        \label{figCDFEliteSucc}
\end{figure}

Figure \ref{figCDFEliteSucc} shows a cumulative distribution function of  $\kappa$ for startups whose founders are elite HEIs alumni. A founder, or a group of founders from the same company, is a product of an elite HEI if the average IGC of all HEIs she/he attended is over or equal to four. There is a positive correlation between mature ecosystems and fundraising capability. In addition, startups whose founders are elite HEI alumni tend to increase $\kappa$.  Finally, the combination of a mature ecosystem and elite HEI affiliation correlates with better fundraising capability, again supporting our hypothesis that elite HEIs have more influence on regional startup ecosystems. 

\section{Discussion}\label{sec:find}

In this study, we investigate how HEIs contribute to Brazil's regional entrepreneur networks, and the nature of these networks. Our original hypothesis suggested that entrepreneur networks in regionally disadvantaged areas, such as the North and Northeast of Brazil, will be closely linked with networks from elite HEIs in the wealthier South and Southeast. Though we find that elite HEIs, such as PUC/SP, UNICAMP, and IBMEC, have the longest spatial ranges, their network influence is still mostly local and for instance does not extend to the North and Northeast. Overall, most of the connections between HEIs and startups are local. The most elite HEI within the region tends to have the most influence within the regional network. We also found a strong presence of Stanford University in our networks, likely reflecting the global influence of this university and Silicon Valley in business education, technology, entrepreneurship, and innovation.


In terms of variability among regional entrepreneur ecosystems, we found that the nature of networks varied by region, given their varied level of development. While IT and Telecom are the most common sectors across regions, there is more variability in startup sectors in the wealthier Southeast than in other areas. In addition, the most economically disadvantaged area, has a strong presence of health startups, likely due to historical underdevelopment. This possibly is an avenue for future research in terms of the potential of startups aimed at alleviating gaps in social service provision in low-income contexts.

Overall, we find support for our hypothesis that regional elite HEIs (the HEI with the highest educational quality rankings in the region) influence regional startup ecosystems and the fundraising capability of founders. We also found that most startup founders were contemporaries while at university, meaning that they overlapped during their course of study, or met through social networks during or after their studies. We found that a majority of startup founders studied computer science, likely reflective of the strong presence of IT and telecom startups in Brazil.

Regarding the limitations of our dataset, though Crunchbase is the most comprehensive data source for startups, it does not include all startups.  A lower percentage of startups are registered in Crunchbase on developing and emerging economies. Second, as we relied on data entered by individual users on LinkedIn to furnish information on founders, there may be entry mistakes or misinformation. However, we believe that this is rare given that founders would be incentivized to invest in a proper online profile. Third, founder characteristics may be biased towards those that are LinkedIn users, and therefore might not be representative. Overall, we used the newest data sources, and are not aware of any existing studies targeting middle-income countries or low-income regions.




\section{Conclusion}\label{sec:con}

In this study, we take a unique conceptual approach in examining the influence of HEIs on entrepreneur networks in Brazil, a middle-income country with low-income regions. We employ an innovative methodological approach by mining publicly available data from Crunchbase, LinkedIn, and the official index of higher education institution quality, to construct and examine the social networks of startup founders. We find that most of the founders were contemporaries at the same HEI and that entrepreneurs frequently seek additional training after startup creation. We observe that the most influential nodes in the network are elite HEIs, though they usually remain within a more localized geographical range. The most nationally prestigious HEIs in the South and Southeast have the longest spatial range into other regions, yet remain fairly local, nor do they extend into the economically disadvantaged North and Northeast. In addition, we find that HEI quality and the maturity of the ecosystem influence startup success. While all of our mature startup ecosystems are in the wealthier South and Southeast, we see some regional movement in the top emerging ecosystems. 

Our findings, therefore, inform research in emerging, developing, and developed countries aiming to stimulate higher education and entrepreneurship particularly in a context of regional inequality. We find support for the notion that contemporaries at HEIs, and in particular, elite HEIs, have powerful influences on entrepreneur social networks. Our findings contribute to education, entrepreneurship, and development research more globally than studies exclusively focused on high-income countries, by examining entrepreneur networks in a middle-income country, Brazil, that also has low-income regions.





\bibliographystyle{splncs04}
\bibliography{refs}

\begin{thebibliography}{10}
\providecommand{\url}[1]{\texttt{#1}}
\providecommand{\urlprefix}{URL }
\providecommand{\doi}[1]{https://doi.org/#1}

\bibitem{Xray2017}
Abstartups: {O momento da startup Brasileira e o futuro do ecossistema de
  inovação}. {Abstartups and Accenture} (2018), {Available online at:
  \url{http://abstartups.com.br/PDF/radiografia-startups-brasileiras.pdf}}

\bibitem{AIEC}
AIEC: {Faculdade AIEC}, \url{{https://www.aiec.br/}}, {Online; accessed
  14-April-2019}

\bibitem{Balestra2018}
Balestra, C., Llena-Nozal, A., Murtin, F., Tosetto, E., Arnaud, B.:
  {Inequalities in emerging economies} (Dec 2018),
  \url{https://doi.org/10.1787/6c0db7fb-en}

\bibitem{BANERJI201946}
{Banerji}, D., {Reimer}, T.: {Startup founders and their LinkedIn connections:
  Are well-connected entrepreneurs more successful?} Computers in Human
  Behavior  \textbf{90},  46--52 (Jan 2019),
  \url{{https://doi.org/10.1016/j.chb.2018.08.033}}

\bibitem{BATES1997109}
Bates, T.: {Financing small business creation: The case of Chinese and Korean
  immigrant entrepreneurs}. Journal of Business Venturing  \textbf{12}(2),
  109--124 (1997), \url{{https://doi.org/10.1016/S0883-9026(96)00054-7}}

\bibitem{BURT2000345}
Burt, R.S.: {The network structure of social capital}. Research in
  Organizational Behavior  \textbf{22},  345--423 (2000),
  \url{{https://doi.org/10.1016/S0191-3085(00)22009-1}}

\bibitem{Comission2013}
{Commision of the European communities}: {Entrepreneurship in Europe} (2013),
  \url{{http://ec.europa.eu/invest-in-research/pdf/download_en/entrepreneurship_europe.pdf}}

\bibitem{CrunchBase}
Crunchbase: Main site, \url{{https://www.crunchbase.com/}}, {Online; accessed
  14-April-2019}

\bibitem{Dalle2017}
{Dalle}, J.M., den {Besten}, M., {Menon}, C.: {Using Crunchbase for economic
  and managerial research} (Nov 2017),
  \url{{https://doi.org/10.1787/6c418d60-en}}

\bibitem{Dasgupta1994}
Dasgupta, P., David, P.: {Toward a new economics of science}. Research Policy
  \textbf{23},  487--521 (1994),
  \url{{https://doi.org/10.1016/0048-7333(94)01002-1}}

\bibitem{ETZKOWITZ2000313}
Etzkowitz, H., Webster, A., Gebhardt, C., Terra, B.R.C.: {The future of the
  university and the university of the future: evolution of ivory tower to
  entrepreneurial paradigm}. Research Policy  \textbf{29}(2),  313--330 (2000),
  \url{{https://doi.org/10.1016/S0048-7333(99)00069-4}}

\bibitem{crunch12}
{Eugene}, L.Y., {Yuan}, S.D.: {Where's the Money? {T}he Social Behavior of
  Investors in Facebook's Small World}. In: {2012 IEEE/ACM International
  Conference on Advances in Social Networks Analysis and Mining}. pp. 158--162
  (Aug 2012), \url{{https://doi.org/10.1109/ASONAM.2012.36}}

\bibitem{CensoHEI}
INEP: {Censo da Educação Superior},
  \url{{http://inep.gov.br/educacao-superior}}, {Online; accessed
  14-April-2019}

\bibitem{IGC}
INEP: {Indice geral de cursos (IGC)},
  \url{{http://inep.gov.br/en/indice-geral-de-cursos-igc-}}, {Online; accessed
  14-April-2019}

\bibitem{INEP}
INEP: {Instituto Nacional de Estudos e Pesquisas Educacionais Anísio
  Teixeira}, \url{{http://portal.inep.gov.br/}}, {Online; accessed
  14-April-2019}

\bibitem{CENSUP}
INEP: {Sistema do censo da educação superior (CENSUP)},
  \url{{http://sistemascensosuperior.inep.gov.br/censosuperior_2018}}, {Online;
  accessed 14-April-2019}

\bibitem{Jaffe1986}
Jaffe, A.B.: {Technological opportunity and spillovers of R\&D: Evidence from
  firms' patents, profits, and market Value}. American Economic Review
  \textbf{76}(5),  984--1001 (Dec 1986), \url{{https://doi.org/10.3386/w1815}}

\bibitem{Gremm2018}
Julia, G., Julia, B., Fietkiewicz, K., Stock, W.: {Transitioning Towards a
  Knowledge Society: Qatar as a Case Study}. Springer (Jan 2018),
  \url{{https://doi.org/10.1007/978-3-319-71195-9}}

\bibitem{Klyver2008}
Klyver, K., Hindle, K., Meyer, D.: {Influence of social network structure on
  entrepreneurship participation—A study of 20 national cultures}, pp.
  331--347. Springer International Publishing, US (Sep 2008),
  \url{{https://doi.org/10.1007/s11365-007-0053-0}}

\bibitem{Kumar2013}
Kumar, K.B., van Welsum, D.: {Knowledge-based economies and basing economies on
  knowledge: Skills a missing link in GCC countries}. RAND Corporation (2013),
  {Available online at:
  \url{https://www.rand.org/pubs/research_reports/RR188.html}}

\bibitem{LARSON1991173}
Larson, A.: {Partner networks: Leveraging external ties to improve
  entrepreneurial performance}. Journal of Business Venturing  \textbf{6}(3),
  173--188 (May 1991), \url{{https://doi.org/10.1016/0883-9026(91)90008-2}}

\bibitem{Lechner2003}
Lechner, C., Dowling, M.: {Firm networks: external relationships as sources for
  the growth and competitiveness of entrepreneurial firms}. Entrepreneurship \&
  Regional Development  \textbf{15}(1),  1--26 (2003),
  \url{https://doi.org/10.1080/08985620210159220}

\bibitem{Light1984}
Light, I.: {Immigrant and ethnic enterprise in North America}. Ethnic and
  Racial Studies  \textbf{7}(2),  195--216 (Sep 1984),
  \url{https://doi.org/10.1080/01419870.1984.9993441}

\bibitem{Lima2012}
Lima, A., Musolesi, M.: {Spatial dissemination metrics for location-based
  social networks}. In: Proceedings of the 2012 ACM Conference on Ubiquitous
  Computing. pp. 972--979 (Sep 2012),
  \url{{https://doi.org/10.1145/2370216.2370429}}

\bibitem{McCowan2007}
{McCowan, T.}: {Expansion without equity: An analysis of current policy on
  access to higher education in Brazil}. {Higher Education}  \textbf{53},
  {579--598} (May 2007), \url{{https://doi.org/10.1007/s10734-005-0097-4}}

\bibitem{Msigwa2016}
{Msigwa, Faustina M.}: {Widening participation in higher education: a social
  justice analysis of student loans in Tanzania}. {Higher Education}
  \textbf{72}(4),  541--556 (Oct 2016),
  \url{{https://doi.org/10.1007/s10734-016-0037-5}}

\bibitem{UNESCO2016}
{Nations Educational, Scientific and Cultural Organization (UNESCO)}:
  {Education 2030: Towards inclusive and equitable quality education and
  lifelong learning for all}. UNESCO ({2016}), {Available online at:
  \url{https://unesdoc.unesco.org/ark:/48223/pf0000245656}}

\bibitem{Newman2010}
Newman, M.: {Networks: An introduction}. Oxford University Press, Inc., New
  York, NY, USA (2010),
  \url{{https://doi.org/10.1093/acprof:oso/9780199206650.001.0001}}

\bibitem{newman2003mixing}
Newman, M.E.: {Mixing patterns in networks}. Physical Review E  \textbf{67}(2),
   026126 (Feb 2003), \url{{https://doi.org/10.1103/PhysRevE.67.026126}}

\bibitem{Nuscheler2016}
Nuscheler, D.: {Regularly change a running system! An analysis of
  stage-specific criteria for attracting venture capital and changing the
  likelihood for getting funded} (2016), {Available online at:
  \url{http://ifabs.org/assets/stores/1206/userfiles/3IFABS\%20Best\%20Poster\%20Award\%20-\%20Daniela\%20Nuscheler,\%20TU\%20Dortmund\%20University,\%20DE.pdf}}

\bibitem{OECD}
OECD: {Rethinking quality assurance for higher education in Brazil} (2018),
  \url{{https://doi.org/https://doi.org/10.1787/9789264309050-en}}

\bibitem{Owen2004}
Owen-Smith, J., Powell, W.: {Knowledge networks as channels and conduits: The
  effects of spillovers in the Boston biotechnology community}. Organization
  Science - ORGAN SCI  \textbf{15},  5--21 (Feb 2004),
  \url{{https://doi.org/10.1287/orsc.1030.0054}}

\bibitem{Perotti2015}
{Perotti, Victor and Yu, Yang}: {Startup Tribes: Social Network Ties that
  Support Success in New Firms}. {AMCIS 2015 Proceedings}  (2015), {Available
  online at:
  \url{http://citeseerx.ist.psu.edu/viewdoc/download?doi=10.1.1.852.4582&rep=rep1&type=pdf}}

\bibitem{porter2005institutional}
Porter, K., Whittington, K.B., Powell, W.W.: {The institutional embeddedness of
  high-tech regions: relational foundations of the Boston biotechnology
  community}. Clusters, networks, and innovation  \textbf{261}, ~296 (2005),
  {Available online at:
  \url{https://web.stanford.edu/group/song/papers/Porter_etal.pdf}}

\bibitem{Renzulli2005}
Renzulli, L.A., Aldrich, H.: {Who can you turn to? Tie activation within core
  business discussion networks}. Social Forces  \textbf{84}(1),  323--341
  (2005), \url{{https://doi.org/10.1353/sof.2005.0122}}

\bibitem{LABRA201678}
{Romilio Labra and Juan Antonio Rock and Isabel Álvarez}: {Identifying the key
  factors of growth in natural resource-driven countries. A look from the
  knowledge-based economy}. {Ensayos sobre Política Económica}
  \textbf{34}(79),  78--89 (Apr 2016),
  \url{{https://doi.org/10.1016/j.espe.2015.12.001}}

\bibitem{Rosenberg1994}
Rosenberg, N., Nelson, R.: {American universities and technical advance in
  industry}. Research Policy  \textbf{23}(3),  323--348 (May 1994),
  \url{{https://doi.org/10.1016/0048-7333(94)90042-6}}

\bibitem{Sanderson2017}
Sanderson, T.: {Education remains the catalyst for Brazil’s staggering
  inequality} (Nov 2017),
  \url{{https://brazilian.report/society/2017/11/06/education-brazil-staggering-inequality/}}

\bibitem{Silva2017}
{Simone Affonso da Silva}: {Regional inequalities in Brazil: Divergent readings
  on their origin and public policy design}. EchoGeo  ({2017}), {Available
  online at: \url{https://journals.openedition.org/echogeo/15060}}

\bibitem{Tata2016}
Tata, A., Laureiro~Martinez, D., Brusoni, S.: {Don’t look back? The effect of
  attention to time and self on startup funding}. Academy of Management
  Proceedings  \textbf{2016}(1),  13926 (Jan 2016),
  \url{{https://doi.org/10.5465/ambpp.2016.13926abstract}}

\bibitem{TATA201738}
Tata, A., Martinez, D.L., Garcia, D., Oesch, A., Brusoni, S.: {The
  psycholinguistics of entrepreneurship}. Journal of Business Venturing
  Insights  \textbf{7},  38--44 (2017),
  \url{{https://doi.org/10.1016/j.jbvi.2017.02.001}}

\bibitem{Rank}
{Times Higher Education}: {The World Universitiy Rankings},
  \url{{https://www.timeshighereducation.com/world-university-rankings/2019/world-ranking#!/page/0/length/25/sort_by/rank/sort_order/asc/cols/stats}},
  {Online; accessed 14-April-2019}

\bibitem{Zimmer1987}
Zimmer, C., Aldrich, H.: {Resource mobilization through ethnic Networks:
  Kinship and friendship ties of shopkeepers in England}. Sociological
  Perspectives  \textbf{30}(4),  422--445 (Oct 1987),
  \url{{https://doi.org/10.2307/1389212}}

\end{thebibliography}
 
\appendix
\newpage
\section{Dataset Overview}

\begin{table}[!tbh]
\centering
\scriptsize
\caption{Dataset overview.}
\begin{tabular}{lr} \hline
\textbf{Startup creation period} & 2004 to 2018
\\ \hline
\textbf{Number of startups} &  1,547 \\ \hline
\textbf{Number of founders} & 454 \\ \hline
\textbf{Number of HEIs}     & 146  \\ \hline
\textbf{Number of degrees obtained by founders} & 648  \\ \hline
\end{tabular}
\label{tab:Overview}
\end{table}

\newpage
\section{Illustration of the Networks Studied}\label{sec:app-nets}


\begin{figure}[!htb]
    \centering
    \includegraphics[width=0.99\textwidth]{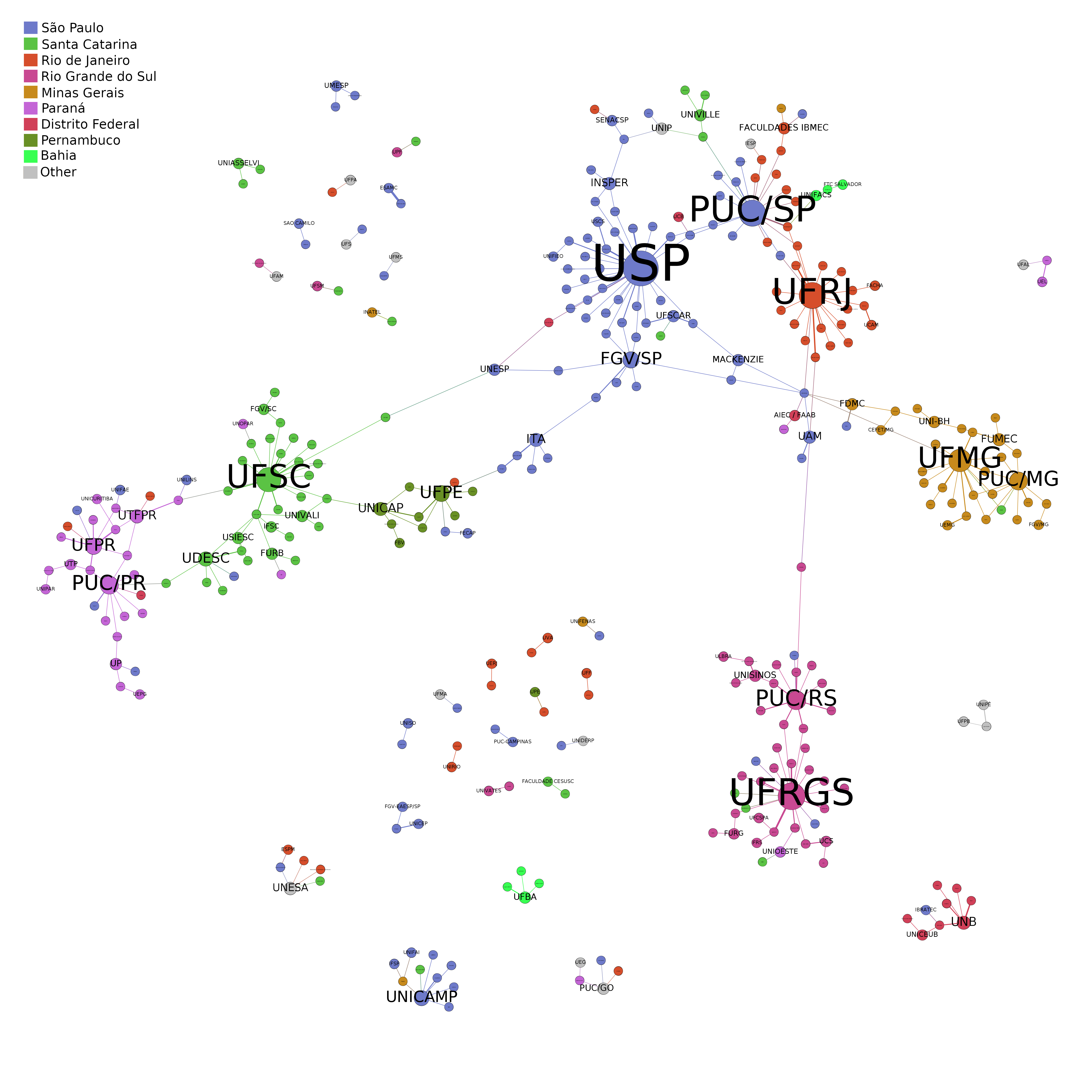}
    \caption{{\tt Undergrad} network. Node colors represent the Brazilian state where they are located in.}
    \label{fig:NetUndergradStartupsAcdemic}
\end{figure}


\begin{figure}[!htb]
    \centering
    \includegraphics[width=0.99\textwidth]{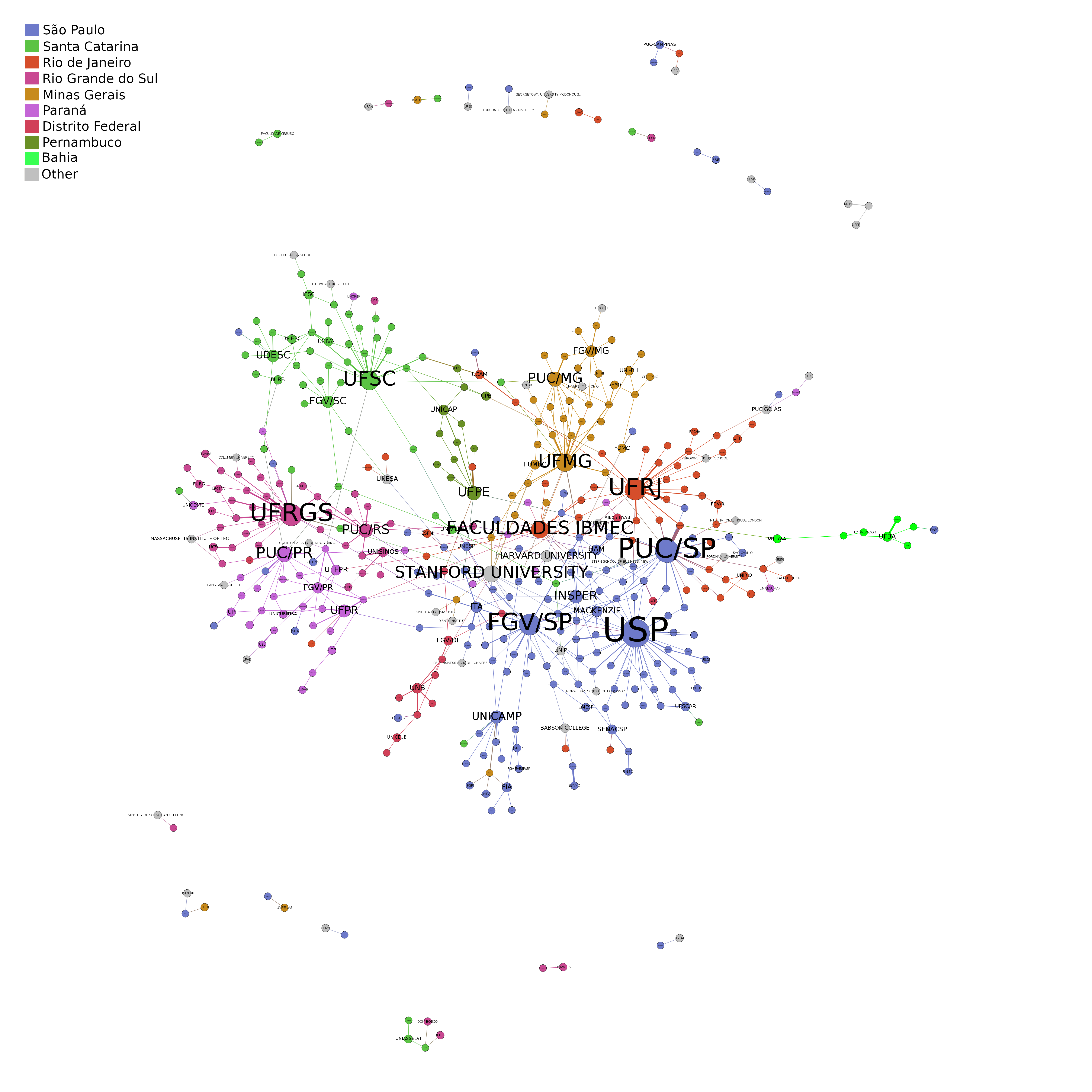}
    \caption{{\tt All-Degree} network. Node colors represent the Brazilian state where they are located in.}
    \label{fig:NetAllStartupsAcdemic}
\end{figure}

\newpage
\clearpage
\section{Full University Names and Abbreviations}\label{sec:app-fullnames}

\begin{itemize}
    \item Universidade de São Paulo (USP) 
    \item Universidade Anhembi Morumbi (UAM)
    \item Unidervsidade Estadual Paulista (UNESP) 
    \item Universidade Presbiteriana Mackenzie (MACKENZIE)
    \item Faculdade Milton Campos (FDMC)
    \item Faculdade de Administração de Brasília (FAAB)
    \item Fundação Getúlio Vargas de São Paulo (FGV/SP)
    \item Instituto de Ensino e Pesquisa (INSPER)
    \item Instituto Brasileiro de Mercado de Capitais (IBMEC)
    \item Associação Internacional de Educação Continuada (AIEC)
    \item Universidade Federal do Minas Gerais (UFMG)
    \item Universidade Federal do Rio de Janeiro (UFRJ)
    \item Universidade Federal do Santa Catarina (UFSC) 
    \item Universidade Federal do Grande do Sul (UFRGS)
    \item Universidade Federal do Estado do Rio de Janeiro (UNIRIO) 
    \item Pontifícia Universidade Católica de Paraná (PUC/PR)
    \item Pontifícia Universidade Católica de São Paulo (PUC/SP)
    \item Pontifícia Universidade Católica de Minas Gerais (PUC/MG)
    \item Pontifícia Universidade Católica de Rio Grande do Sul (PUC/RS)
\end{itemize}

\newpage
\section{Startup Sectors by Region}\label{sec:app-region}

\begin{figure}[!tbh]
    \centering
    \begin{minipage}{0.45\textwidth}
        \centering
        \includegraphics[width=0.99\textwidth]{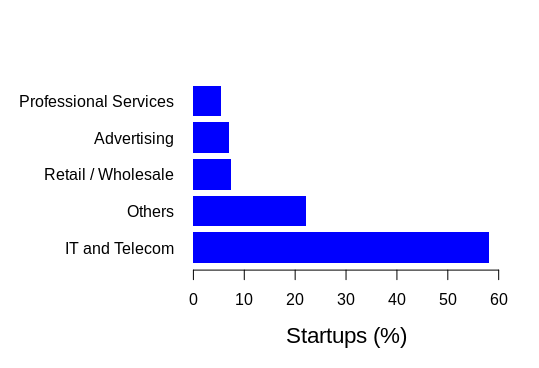}
        \caption{South.}
        \label{fig:categoriesStartups-South}
    \end{minipage}
    \hfil
    \begin{minipage}{.45\textwidth}
        \centering
        \includegraphics[width=0.99\textwidth]{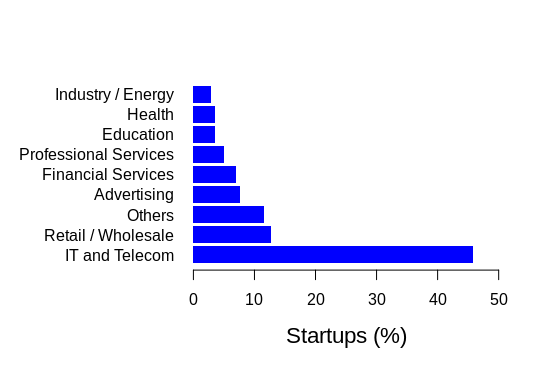}
        \caption{Southeast.}
        \label{fig:categoriesStartups-Southeast}
    \end{minipage}
\end{figure}

\begin{figure}[!tbh]
    \centering
    \begin{minipage}{0.45\textwidth}
        \centering
        \includegraphics[width=0.99\textwidth]{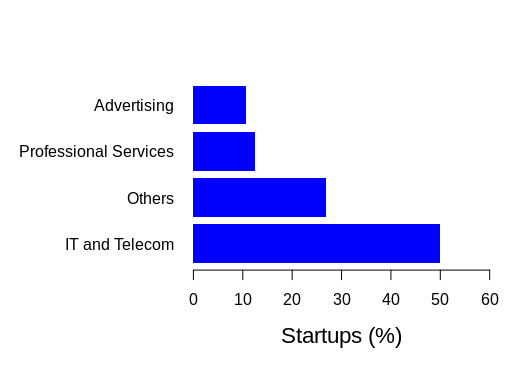}
        \caption{Central-West.}
        \label{fig:categoriesStartups-West-Centre} 
    \end{minipage}
    \hfil
    \begin{minipage}{.45\textwidth}
        \centering
        \includegraphics[width=0.99\textwidth]{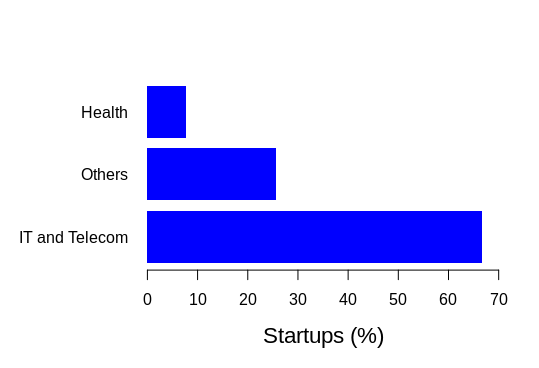}
        \caption{Northeast.}
        \label{fig:categoriesStartups-Northeast}
    \end{minipage}
\end{figure}

\newpage
\section{Summary of All Ecosystems Studied}\label{sec:app-eco}

\begin{table}[h]
\scriptsize
\centering
\caption{Summary of all ecosystems studied.}

\begin{tabular}{l l l r r l l l r r}
\hline
\textbf{Ecosystem}  & \textbf{State} & \textbf{Region} & \textbf{\# of startups} & \textbf{Fundraising}  \\ 
\hline
São Paulo           & SP & Southeast    & 458     & \$2.6B   \\
Rio de Janeiro      & RJ & Southeast    & 323     & \$283.6M \\
Belo Horizonte      & MG & Southeast    & 151     & \$33.8M  \\
Porto Alegre        & RS & South        & 114     & \$11.1M  \\
Curitiba            & PR & South        & 95      & \$67.3M  \\
Florianópolis       & SC & South        & 94      & \$43.5M  \\
Brasilia            & DF & Central-West & 39      & \$3.7M   \\
Recife              & PE & Northeast    & 38      & \$8.5M   \\
Campinas            & SP & Southeast    & 36      & \$13.4M  \\
Fortaleza           & CE & Northeast    & 28      & \$105.4K \\
São José dos Campos & SP & Southeast    & 25      & \$4.1M   \\
Goiânia             & GO & Central-West & 22      & no info. \\
Barueri             & SP & Southeast    & 21      & \$10.9M  \\
Joinville           & SC & South        & 21      & \$41.3M  \\
Uberlândia          & MG & Southeast    & 15      & \$2.6M   \\
João Pessoa         & PB & Northeast    & 14      & \$194.8K \\
Itá                 & PB & Northeast    & 11      & \$2.8M   \\
Jussara             & BA & Northeast    & 11      & \$130.1K \\
Salvador            & BA & Northeast    & 11      & \$1.8M   \\
Blumenau            & SC & South        & 10      & \$1.3M   \\
Sorocaba            & SP & Southeast    & 10      & \$530.3K \\

\end{tabular}
\end{table}

\end{document}